\documentclass[%
  twocolumn,
 showpacs,
 showkeys,
 preprintnumbers,
 amsmath,amssymb,
 aps,
  pra,
  longbibliography,
 ]{revtex4-1}

\usepackage{amssymb,amsthm,amsmath}

\usepackage{tikz}
\usepackage[breaklinks=true,colorlinks=true,anchorcolor=blue,citecolor=blue,filecolor=blue,menucolor=blue,pagecolor=blue,urlcolor=blue,linkcolor=blue]{hyperref}
\usepackage{graphicx}
\usepackage{url}

\usepackage{xcolor}

\begin{document}

\title{Space and time in a quantized world}


\author{Karl Svozil}
\affiliation{Institute for Theoretical Physics, Vienna
    University of Technology, Wiedner Hauptstra\ss e 8-10/136, A-1040
    Vienna, Austria}
\email{svozil@tuwien.ac.at} \homepage[]{http://tph.tuwien.ac.at/~svozil}

\pacs{03.65.Ta, 03.65.Ud}
\keywords{construction of space-time, relativity theory, quantum mechanics, conventions, entanglement, signal propagation, field theory}

\begin{abstract}
Rather than an {\em a priori} arena in which events take place, space-time is a construction of our mind making possible a particular kind of ordering of events. As quantum entanglement is a property of states independent of classical distances, the notion of space and time has to be revised to represent the holistic interconnection of quanta. We also speculate about various forms of reprogramming, or reconfiguring, the propagation of information for multipartite statistics and in quantum field theory.
\end{abstract}

\maketitle

\section{Intrinsic construction of space-time frames}

Physical space and time appear to be {\em ordering events} by quantifying top-bottom, left-right, front-back, as well as before-after.
In that function, space-time relates to actual physical events, such as clicks in particle detectors.
Without such events, space-time would be metaphysical at best, because there would be no operational basis
that gave meaning to the aforementioned categories.
Intrinsic space-time is tied to, or rather based upon, physical events;
and is bound to operational means available to observers ``located inside'' the physical system.

In acknowledging this empirical foundation,
Einstein's centennial paper on space-time \cite{ein-05,naber},
and to a certain extent Poincare's thoughts \cite{poincare02}, introduced
{\em conventions and operational algorithmic procedures} that allow
the generation of space-time frames by relying on intrinsically feasible methods and techniques alone \cite{toffoli:79,svozil-94}.
This renders a space-time (in terms of clocks, scales and conventions for
the definition of space-time frames, as well as their transformations) which
is {\em means relative} \cite{Myrvold2011237} with respect to physical devices (such as clocks and scales), as well as
to procedures and conventions
(such as for {\em defining} simultaneity employing round-trip time, which is
nowadays even used by {\em Cristian's Algorithm} for computer networks).
These unanimously executable measurements and ``algorithmic'' physical procedures
need not rely upon any kind of absolute metaphysical knowledge (such as ``absolute space or time'').

This approach is characterized by constructing operational, intrinsic space-time frames based on physical events alone;
rather then by staging physical events in a Kantian {\it a priori} ``space-time theatre.''
One step in this direction is, for instance, the determination of the {\em dimensionality} of space and of space-time from
empirical evidence \cite{sv2}.

Consequently, space and time emerge as concepts that are not independent of the physical phenomena
(as well as on assumptions or conventions) by which they are constructed.
Therefore, it is quite legitimate to ask whether the space-time of classical physics can be
carried over to quantum space-time \cite{Kreinovich-94,Myrvold2002435}.

\section{Encoding information on single quanta}

So far, there is evidence that any kind of ``will- and useful'' classical or quantum
information in terms of nonrandom bit(stream)s  can be transferred from some space-time point $A$
to another space-time point $B$ only {\em via} individual quanta:
single quanta are emitted at some space-time point   $A$, and absorbed at another space-time point $B$.
This is true, in particular, for quantum teleportation; that is, the entanglement assisted transmission of quantum information from one location to another.
Thus we shall first concentrate on the generation of quantum space time
-- that is, on the construction of clocks and scales based upon quantum processes
yielding space-time frames, as well as on their transformations by  of a direct bit exchange.
Issues often referred to as quantum ``nonlocality'' and entanglement are relegated to the next section.

\subsection{Time scales}

If indeed one takes seriously the idea that ``quanta can be utilized to create space-time frames,''
then we need to base space and time scales used in such frames on quantum mechanical entities,
that is, on quantum clocks and on quantum scales.

Formally, by Cayley's representation theorem the unitary quantum evolution can be represented by some subgroup of the symmetric group.
One approach to quantum clocks and time might thus be to consider general distances and metrics on permutations, in particular,
on the symmetric groups, thereby relating changes in quantum states to time.

Indeed, the current definition of the second in the International System (SI) of units
is {\it via} 9 192 631 770 transitions between two orthogonal quantum states of a caesium 133 atom.
That is, if we encode the two ground states by the subspaces spanned by the two orthogonal vectors
$\vert \psi_0 \rangle \equiv (0,1)$ and $\vert \psi_1 \rangle \equiv (1,0)$,
[or, equivalently, by the projectors $\text{diag}(0,1)$    and  $\text{diag}(1,0)$]
in two-dimensional Hilbert space,
then the 9 192 631 770'th fraction of a second is delivered by the unitary operator that is known as the
{\em not gate} \cite{mermin-07} $\textsf{\textbf{X}}
=   \begin{pmatrix}0& 1\\1&0
\end{pmatrix}
$,
representing a single permutation-transition $\textsf{\textbf{X}}\vert \psi_i \rangle $  between $\vert \psi_i \rangle  \leftrightarrow \vert \psi_{i\oplus 1} \rangle $,
$i\in \{ 0, 1\}$,
of two orthogonal quantum states of a caesium 133 atom.

\subsection{Space scales}

The current definition of spatial distances in the International System of units
is in terms of the propagation of light quanta in vacuum.
More specifically, the metre is the length of the path travelled by light in vacuum during a time interval
of 1/299 792 458'th part of a second
--
or, equivalently,
as light travels 299 792 458 metres per second,
a duration in which 9 192 631 770 transitions between two orthogonal quantum states of a caesium 133 atom occur
--
during
9 192 631 770/299 792 458 $\approx 31$ transitions of two orthogonal quantum states of a caesium 133 atom.

More generally we may ask
what, exactly, is a {\em ``spatial distance?''}
In particular, what quantum meaning can be ascribed to a ``path travelled by light in vacuum?''
First and foremost, any spatial distance seems to depend on two criteria:
(i) separateness, or disconnectedness; as well as
(ii) the capacity to (inter-)connect.
The latter connection must, by quantum rules,
be mediated {\it via} {\em permutations.}
In the simplest sense, one could algorithmically model such a contact transmission by reversible {\em cellular automata}
\cite{fredkin,svozil-1996-time,thooft-2013}; that is, by a tesselated, three-dimensional, discrete computation space \cite{zuse-67}
constantly permuting itself.

\subsection{Alexandrov-Zeemann theorem}

In order to make operational sense without regress to absolute space-time frames,
the SI definition of length implicitly assumes that the velocity of light in vacuum for all space-time frames is constant,
regardless of the state of motion of that frame
\cite{peres-84}.
By these assumptions and other conventions, such as Einstein's definition of simultaneity \cite{ein-05} and bijectivity of coordinate transformations,
the Lorentz transformations are essentially (up to shift-translations and
dilations with positive scalar constants) a consequence
of the Alexandrov-Zeemann theorem of incidence geometry \cite{alex3,zeeman,lester,naber}.
Pointedly stated, if two observers ``presiding over their reference frames agree'' \cite{naber} that points connected by
light rays can be interconnected, then linear transformations of space-time frames follow.

From a purely formal point of view,
fixing the invariance (constancy) of the velocity of light with respect to changes of space-time frames
appears to be purely conventional,
and thus may be even considered as arbitrary and {\it a priori} unjustified, if not misleading.
Any other velocity, both sub- as well as superluminal
-- even associated with no-signalling correlated events such as from phased arrays (see below) --
would suffice for the construction of transformations between space-time frames.

The physical motivation for choosing light in vacuum is twofold:
First, the  {\em form invariance}
of the equations of motion, such as Maxwell's equation in vacuum, is a convenience.
And secondly,
all space-time frames correctly reflect the causality relative to the electromagnetic interaction.

It is thus suggested to ``stay within a single type of interaction'' when it comes to the construction of clocks and scales,
and also to fix the invariance of the respective signals for the construction of space-time frames, as well as the transformation laws between them.
The resulting space-time is defined means relative to (the causality induced by) this interaction \cite{svozil-relrel}.
In this sense, the SI definition renders a space-time with is means relative to the electromagnetism.

\section{Encoding information across quanta}

\subsection{Entanglement characteristics}

At the time of conceptualizing special relativity theory, quantum mechanics was in its infancy,
and quantum effects were therefore not considered for the definition of space-time scales.
Alas, this has changed since Schr\"odinger pointed out the possibility of entangled quantum states of multipartite quantized systems;
states that do not have any classical local counterpart.
Entanglement is characterized by an encoding of (classical) information ``across quanta'' \cite{zeil-99,zeil-Zuk-bruk-01,svozil-2002-statepart-prl}
that defy any kind of spatial apartness or locality,
and yield experimental violations \cite{wjswz-98} of classical probabilities \cite{pitowsky}.
These features alone suggest to reconsider quantum mechanical processes for the definition of space-time frames.

One of the characteristics of quantum entanglement is that information is not encoded in the single quanta which constitute
an entangled system. Therefore, through context translation, any enquiry about the state of a single quantum
is futile, because no such information is available prior to this ``forced measurement.''
The archetypical example of this situation is the Bell state
$\vert \Psi_- \rangle = \left( 1/\sqrt{2} \right) \left(\vert +-\rangle - \vert -+\rangle \right)$.
On the one hand, $\vert \Psi_- \rangle$ is totally and irreducible indeterminate about the states $\vert -\rangle$ or $\vert +\rangle$
of its individual two constituents.
Indeed a ``forced measurement'' yields random outcomes \cite{svozil-qct}; and the concatenation of independent outcomes
encoded as a binary sequence can, for instance, be expected to be Borel normal
\cite{svozil-2006-ran,PhysRevA.82.022102}; in particular, there is a 50:50 chance for  $\vert -\rangle$ and $\vert +\rangle$, respectively.
On the other hand, $\vert \Psi_- \rangle$ is totally determined by the joint correlations of the particles involved;
in particular, by the two propositions  {\em ``the spin states of the two particles along two orthogonal spatial directions
are different''}   \cite{Zeilinger-97,zeil-99,svozil-2002-statepart-prl}.

Alas, in this view,
for the Bell state as well as for other nonlocalized multipartite entangled states,
in which the constituents can be thought of as ``torn apart'' arbitrary spatial distances,
there is no ``spooky action at a distance'' \cite{Nikolic} whatsoever,
because the multiple constituents, if they become separated and ``drift away'' from their joint space-time preparation regions,
do so at speeds not exceeding the velocity of light; with no further communication or information exchange between them.

Thereby, any greater-than-classical correlations and expectations these constituents carry are due to the particular type
of quantum probabilities.
Recall that the quantum probabilities are generalizations of classical probabilities:
Due to Gleason's theorem the Born rule can be derived from
the noncontextual pasting of blocks of subalgebras (that is, maximal, co-measurable observables);
whereas all classical probability distributions result from convex sums of two-valued states on the Boolean algebra of classical propositions.

Pointedly stated, the so-called ``quantum nonlocality'' is not non-local at all, because
these correlations reside in the (entangled) quantum states which must be perceived holistically (as being one compound state)
rather than as being constructed from separate single quantum states; regardless of the spatial separation
of the constituent quanta forming such states.
The measurements in spatially different regions (regardless of whether they are space-like separated or not)
just recover this property encoded in the quantum states;
thereby nothing needs to be exchanged, nor can information be gained in excess
of the one encoded by the state preparation.

There exist even  quasi-classical models
(which are nonlocal as they require the exchange of one bit per particle pair)
capable of realizing stronger-than-quantum correlations \cite{svozil-2004-brainteaser}.
Claims that these larger-than-classical correlations expresses some kind of
``spooky action at a distance'' mistake correlation for causality.
In this regard, the terminology ``peaceful coexistence'' \cite{shimony-78}
between quantum theory and special relativity, suggesting or even
implying some perceivable kind of inconsistency between them,
is misleading, because there cannot occur any kind of ``clash'' or inconsistency between fundamental observables and processes and any entities, such as space-time, which are secondary constructions of the mind, based on the former observables and processes.

\subsection{Quantum statistics}

The remaining discussion is very speculative and should not be taken as claiming the existence of any faster-than-light signalling.

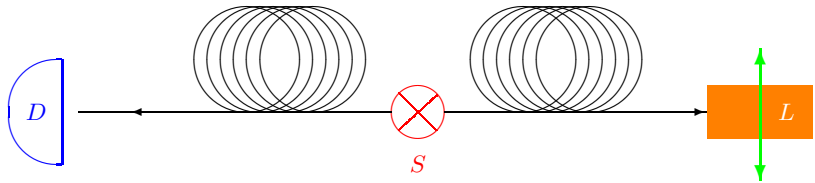
\begin{figure*}
\begin{center}
\unitlength 0.7mm 
\linethickness{0.4pt}
\ifx\plotpoint\undefined\newsavebox{\plotpoint}\fi 
\begin{picture}(175,50)(20,0)
\put(100,20){\color{red}\circle{10}}
\put(105,20){\line(1,0){50}}
\put(95,20){\line(-1,0){50}}
\put(120,30){\circle{25}}
\put(80,30){\circle{25}}
\put(122.5,30){\circle{25}}
\put(77.5,30){\circle{25}}
\put(125,30){\circle{25}}
\put(75,30){\circle{25}}
\put(127.5,30){\circle{25}}
\put(72.5,30){\circle{25}}
\put(130,30){\circle{25}}
\put(70,30){\circle{25}}
\put(132.5,30){\circle{25}}
\put(67.5,30){\circle{25}}
\put(32.25,20){\color{blue}\oval(20,20)[l]}
\put(32.5,10){\color{blue}\line(0,1){20}}
\put(27.5,20){\makebox(0,0)[cc]{\color{blue}$D$}}
\put(45.5,20){\vector(-1,0){.07}}\put(55.5,20){\line(-1,0){20}}
\put(154.5,20){\vector(1,0){.07}}\put(144.5,20){\line(1,0){20}}
\multiput(96.388,16.608)(.036812183,.033614213){197}{\color{red}\line(1,0){.036812183}}
\multiput(103.503,16.44)(-.033615023,.033615023){213}{\color{red}\line(0,1){.033615023}}
\put(100.056,10.076){\color{red}\makebox(0,0)[cc]{$S$}}
\put(155,15){\color{orange}\rule{20\unitlength}{10\unitlength}}
\put(170,20){\makebox(0,0)[cc]{\color{white}$L$}}
{\thicklines
\put(165,32){\color{green}\vector(0,1){.07}}\put(165,7){\color{green}\vector(0,-1){.07}}\put(165,7){\color{green}\line(0,1){25}}
}
\end{picture}
\end{center}
\caption{(Color online) Speculative delayed choice experiment evoking stimulated emmission-absorption of a quantum constituent in an entangled state.
A singlet state of two quanta is created at source $S$. One of the particle impinges on a detector $D$, the other in a ``box region'' $L$ filled with certain attainable quantum states.}
\label{2013-st1-dcsea}
\end{figure*}

Suppose the constituent quanta of an entangled state are subjected to
{\em active} stimulation rather than passive measurement.
In particular, multi-partite quantum statistics can give rise to stimulated emission or absorption.
For the sake of an attack \cite{svozil-slash} on local causality, consider the delayed choice of, say, either scattering a photon into a ``box of identical photons''
(or directing an electron into a region filled with other electrons occupying certain states attainable by the original electron), or
passing this region without any other identical quanta, as depicted in Fig.\ref{2013-st1-dcsea}.
One might speculate that such a device might be used to communicate a message across the particle pair through controlling the outcome on one side,
thereby {\em spoiling outcome independence,}
because if some agent has free will to ``induce''
some state of one photon of a photon pair in an entangled singlet state,
the other photon has no (random) choice any longer but to scatter into the corresponding state.
One interesting way to argue against such a scenario is by pretending that the source ``(en)forces'' certain statistical properties of the single constituent particles
-- in particular their stochastic behaviour -- of an entangled state even beyond the standard quantum predictions \cite{zeil-99,svozil-2013-omelette}.

Another possibility would be to transmit information across spatially extended quantum states of a large number of particles by affecting
the statistical constraints on one side and observing the effects on the other end.
For the sake of a concrete example consider a superconducting rod which is heated into the nonsuperconducting state (or otherwise ``destroying it'') on one end of the rod,
and observing the gap energy on the other end.

We will turn our attention now to ``second quantization'' effects on single (nonentangled) quanta; in particular, with regard to propagation.
They are due to the presence of (spontaneous or controlled) many-partite excitations of the quantized fields involved.

\section{Field theoretic models of signal propagation}

When considering the propagation of light and other potential signals in vacuum \cite{einstein-aether,dirac-aether},
which will be considered as a {\em signal carrier},
there appear to exist at least two alternative conceptions.
First, we could assume that light is ``attenuated'' by polarization
and other (e.g., quantum statistical) effects.
Without any such interactions such signals might travel arbitrarily fast.
Thus, in order to increase signalling speeds,
we must attempt to disentangle the signal from interacting with the vacuum.
A somewhat related scenario is the hypothetical possibility to ``shift gear'' to another,
less retarding, mode of propagation by (locally) changing the state of the signal carrier; for instance by supercavitation.

A second, entirely different, viewpoint may be that light needs a carrier for propagation;
very much like a phonon needs, or rather subsumes,
collective excitations of some carrier medium.
In such scenarios, stronger couplings might result in higher signalling speeds.

If any such speculation will eventually yield superluminal communication and space travel is highly uncertain,
but should not be outrightly excluded for the mere sake of orthodoxy.
In what follows we briefly mention some possible directions of looking into these issues.

\subsection{Multiple side hopping}

The capacity to transfer information can be modelled by some sort of interconnection between different spatial regions.
One such microphysical model is the vibrating (linear) chain \cite[Sec.~1.2]{Henley-Thirring-EQFT}
which requires some coupled (linearized) oscillators.
The spatial signal carrier is modelled by an interconnected array of coupled oscillators.
Thereby, (the energy of) an excitation is transferred from one oscillator to the next by the coupling between the two.

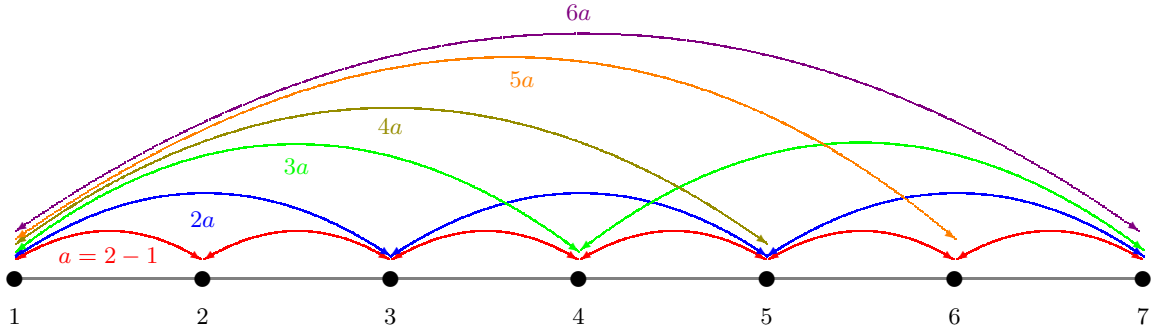
\begin{figure*}
\begin{center}
\unitlength 0.5mm
\linethickness{0.4pt}
\ifx\plotpoint\undefined\newsavebox{\plotpoint}\fi 
\begin{picture}(301,127.817)(0,0)
{\thicklines
\put(0,10){\color{gray}\line(1,0){50}}
\put(50,10){\color{gray}\line(1,0){50}}
\put(99.962,10){\color{gray}\line(1,0){50}}
\put(150,10){\color{gray}\line(1,0){50}}
\put(200,10){\color{gray}\line(1,0){50}}
\put(250,10){\color{gray}\line(1,0){50}}
}
\put(0,10){\circle*{4}}
\put(50,10){\circle*{4}}
\put(100,10){\circle*{4}}
\put(150,10){\circle*{4}}
\put(200,10){\circle*{4}}
\put(250,10){\circle*{4}}
\put(300,10){\circle*{4}}
\thinlines
{
{\put(49.613,15.136){\color{red}\vector(3,-2){.07}}\put(.42,15.136){\color{red}\vector(-3,-2){.07}}\color{red}\qbezier(.42,15.136)(25.437,30.272)(49.613,15.136)}
\put(99.647,15.136){\color{red}\vector(3,-2){.07}}\put(50.454,15.136){\color{red}\vector(-3,-2){.07}}\color{red}\qbezier(50.454,15.136)(75.471,30.272)(99.647,15.136)
\put(149.68,15.136){\color{red}\vector(3,-2){.07}}\put(100.488,15.136){\color{red}\vector(-3,-2){.07}}\color{red}\qbezier(100.488,15.136)(125.505,30.272)(149.68,15.136)
\put(199.714,15.136){\color{red}\vector(3,-2){.07}}\put(150.521,15.136){\color{red}\vector(-3,-2){.07}}\color{red}\qbezier(150.521,15.136)(175.538,30.272)(199.714,15.136)
\put(249.748,15.136){\color{red}\vector(3,-2){.07}}\put(200.555,15.136){\color{red}\vector(-3,-2){.07}}\color{red}\qbezier(200.555,15.136)(225.572,30.272)(249.748,15.136)
\put(299.781,15.136){\color{red}\vector(3,-2){.07}}\put(250.589,15.136){\color{red}\vector(-3,-2){.07}}\color{red}\qbezier(250.589,15.136)(275.605,30.272)(299.781,15.136)}
{
\put(100.067,15.977){\color{blue}\vector(3,-2){.07}}\put(0,15.977){\color{blue}\vector(-3,-2){.07}}\color{blue}\qbezier(0.42,15.977)(50.875,49.613)(100.067,15.977)
\put(200.135,15.977){\color{blue}\vector(3,-2){.07}}\put(100.067,15.977){\color{blue}\vector(-3,-2){.07}}\color{blue}\qbezier(100.067,15.977)(150.942,49.613)(200.135,15.977)
\put(300.202,15.977){\color{blue}\vector(3,-2){.07}}\put(200.135,15.977){\color{blue}\vector(-3,-2){.07}}\color{blue}\qbezier(200.135,15.977)(251.009,49.613)(300.202,15.977)}
{
\put(150.101,17.238){\color{green}\vector(4,-3){.07}}\put(.42,17.238){\color{green}\vector(-4,-3){.07}}\color{green}\qbezier(.42,17.238)(74.42,74.42)(150.101,17.238)
\put(300.202,17.659){\color{green}\vector(4,-3){.07}}\put(150.521,17.659){\color{green}\vector(-4,-3){.07}}\color{green}\qbezier(150.521,17.659)(224.521,74.84)(300.202,17.659)}
{
\put(200.135,19.341){\color{olive}\vector(4,-3){.07}}\put(.42,19.341){\color{olive}\vector(-4,-3){.07}}\color{olive}\qbezier(.42,19.341)(99.857,91.658)(200.135,19.341)}
{
\put(250.168,20.602){\color{orange}\vector(4,-3){.07}}\put(.42,20.602){\color{orange}\vector(-3,-2){.07}}\color{orange}\qbezier(.42,20.602)(137.908,117.306)(250.168,20.602)}
{
\put(298.94,22.704){\color{violet}\vector(4,-3){.07}}\put(.42,22.704){\color{violet}\vector(-3,-2){.07}}\color{violet}\qbezier(.42,22.704)(152.203,127.817)(298.94,22.704)}
\put(0,0){\makebox(0,0)[cc]{$1$}}
\put(50,0){\makebox(0,0)[cc]{$2$}}
\put(100,0){\makebox(0,0)[cc]{$3$}}
\put(150,0){\makebox(0,0)[cc]{$4$}}
\put(200,0){\makebox(0,0)[cc]{$5$}}
\put(250,0){\makebox(0,0)[cc]{$6$}}
\put(300,0){\makebox(0,0)[cc]{$7$}}
\put(25,16){\makebox(0,0)[cc]{\color{red}$a=2-1$}}
\put(50,26){\makebox(0,0)[cc]{\color{blue}$2a$}}
\put(75,40){\makebox(0,0)[cc]{\color{green}$3a$}}
\put(100,50.5){\makebox(0,0)[cc]{\color{olive}$4a$}}
\put(135,63){\makebox(0,0)[cc]{\color{orange}$5a$}}
\put(150,81){\makebox(0,0)[cc]{\color{violet}$6a$}}
\end{picture}
\end{center}
\caption{(Color online) Speculative multiple side hopping might give rise to higher harmonics of the speed of light.}
\label{2013-st1-msh}
\end{figure*}

One possibility to change the resulting signal velocity would be to assume that any oscillator is coupled not only to its next neighbour,
but to other oscillators which are farther apart but nevertheless topologically interconnected. In this way, by increasing the ``hopping distance,''
say, in a periodic medium, as depicted in Fig.~\ref{2013-st1-msh},
faster modes of propagation (as compared to single side hopping) seem conceivable.

We suggest to employ {\em  phased array} (radar) with faster-than-light synchronization, such as the
one enumerated in Table~\ref{2013-tablest1-msh}, of electrical signals
for the exploration of multiple side hopping and the resulting higher order harmonics $2c, 3c,\ldots $ of the velocity of light $c$.
Thereby, the signals generated by the phased array of electrical charges
might resonate with the propagation modes of the field carrying those collective excitations.
For random hopping distances, any such discretization cannot be expected.

In that way, one is not approaching any (supposedly impenetrable) speed-of-light barrier ``from below'' (i.e., with subluminal speeds) but
attempts to induce carrier excitations at almost arbitrary velocities.
We emphasize that the issue of whether or not the vacuum can actually carry such signals is a highly speculative suggestion
that outrightly contradicts long-held beliefs, but remains empirically undecided and unknown.

\begin{table}
\begin{center}
\begin{tabular}{c|ccccccccccc}
\hline\hline
array site&1&2&3&4&5&6&7&$\cdots$\\
{\it versus} time&&&&&&&&\\
\hline
\color{red}
c=1
&\color{red}1&0&0&0&0&0&0&$\cdots$\\
&0&\color{red}1&0&0&0&0&0&$\cdots$\\
&0&0&\color{red}1&0&0&0&0&$\cdots$\\
&0&0&0&\color{red}1&0&0&0&$\cdots$\\
&0&0&0&0&\color{red}1&0&0&$\cdots$\\
&0&0&0&0&0&\color{red}1&0&$\cdots$\\
&0&0&0&0&0&0&\color{red}1&$\cdots$\\
&\multicolumn{8}{l}{$\cdots$}\\
\hline
\color{blue}
c=2
&\color{blue}1&0&0&0&0&0&0&$\cdots$\\
&0&0&\color{blue}1&0&0&0&0&$\cdots$\\
&0&0&0&0&\color{blue}1&0&0&$\cdots$\\
&0&0&0&0&0&0&\color{blue}1&$\cdots$\\
&\multicolumn{8}{l}{$\cdots$}\\
\hline
\color{green}
c=3
&\color{green}1&0&0&0&0&0&0&$\cdots$\\
&0&0&0&\color{green}1&0&0&0&$\cdots$\\
&0&0&0&0&0&0&\color{green}1&$\cdots$\\
&\multicolumn{8}{l}{$\cdots$}\\
\hline
\color{olive}
c=4
&\color{olive}1&0&0&0&0&0&0&$\cdots$\\
&0&0&0&0&\color{olive}1&0&0&$\cdots$\\
&\multicolumn{8}{l}{$\cdots$}\\
\hline
\color{orange}
c=5
&\color{orange}1&0&0&0&0&0&0&$\cdots$\\
&0&0&0&0&0&\color{orange}1&0&$\cdots$\\
&\multicolumn{8}{l}{$\cdots$}\\
\hline
\color{violet}
c=6
&\color{violet}1&0&0&0&0&0&0&$\cdots$\\
&0&0&0&0&0&0&\color{violet}1&$\cdots$\\
&\multicolumn{8}{l}{$\cdots$}\\
\hline\hline
\end{tabular}
\end{center}
\caption{(Color online) Array synchronization for speculative multiple side hopping.
In this discrete setup, the fundamental time unit is the time it takes for (the slowest) light signal in vacuum to propagate (``hop'') one fundamental spatial unit $a$.
}
\label{2013-tablest1-msh}
\end{table}

\subsection{Change of vacuum}

Another possibility to change the propagation velocity of the signal carrier would be to alter its ability
to carry a signal through attenuation and amplification of the processes responsible for sinalling.
The most direct form would be to change the coupling between oscillators in the vibrating chain scheme mentioned earlier.

Another possibility would be to again use quantum statistical effects to reduce or increase the polarizability of the vacuum
by placing bosons or fermions along the signalling path.
A photon, for instance, seems to become accelerated if polarizability is reduced \cite{Scharnhorst-1998,svozil-putz-sol}.

\section{Dimensionality}

One could speculate that the apparent three-dimensionality of physical configuration space is
a reflection of the {\em three-dimensional interconnection} of the signal carrier of this universe
on a very fundamental level.
In this way, information is ``permuted by point contact from one node to the other.''
A discrete version of this would be a three-dimensional cellular automaton.

In another scenario the intrinsic, operational three-dimensionality
is a (fractal) ``shadow'' on a higher dimensional signal carrier \cite{sv4}.
In this view, if there is no ``bending (yielding nontrivial topologies), folding or compactification'' of the extra dimensions involved,
information transfer might become even ``slower'' than in the lower dimensional case, since every extra dimension is nothing but
an extra degree of freedom the bit can pursue, thereby even ``getting lost'' if, say, it travels a direction orthogonal to,
or in other ways inaccessible for, physical three-space.
On the other hand, if this fractal shadow constituting our accessible configuration space
can be bent or even intersected by itself in topologically nontrivial ways,
then information transfer, and thus signalling and space travel, from any point $A$ to any other point $B$
could in principle be obtained with arbitrary velocities.

\section{Concluding remarks}

The short answer of the question of whether quantum space-time is different from classical space-time is this:
since, according to the Alexandrov-Zeemann theorem, bijective space-time transformations are essentially
determined by the {\em causal ordering of events}, any difference of classical {\it versus} quantum space-time
can be reduced to the question of whether or not quantum events can be causally ordered differently than classical ones.
Until now there is not the slightest indication that this is the case,
so there is no evidence of any difference between classical and quantum space-time.
However, there are {\it caveats} to this answer: certain processes,
such as the ones discussed earlier, may give rise to a different quantum ordering,
and thus to different space-times.

With respect to considerations regarding space-time as a construction based on empirical events,
any attempt to unify gravity  as a ``geometrodynamic theory of curved space-time''  {\it on a par} with the standard quantum field theories
must inevitably fail:
if space and time emerge as secondary ``ordering''
concepts based on our primary experience of quanta (e.g. detector clicks),
they cannot be treated on an equal footing with these phenomena.
Thus, if the equivalence principle ``equating'' inertial with gravitational mass is correct,
one could speculate that
the resulting geometrodynamic theory of gravity needs to be based upon some field theoretic effects accounting for this equivalence;
such as ``metrical elasticity'' through vacuum quantum fluctuations \cite{Sakharov-67}.

Beyond electromagnetic and gravitational interactions,
other ``fundamental'' (strong, weak) interactions have been discovered, which, according to the standard unification model,
propagate at the same speed as light, although no direct empirical evidence is available.
In any case, {\it a priori}, different interactions need not always propagate with the same velocity,
making necessary a sort of ``relativized relativity'' \cite{svozil-relrel} that has to cope with
consistency issues, such as the ``grandfather paradox.''
The latter one is also resolved in ``quantum time travelling'' scenarios \cite{svozil-greenberger-2005}.

Insofar multipartite and field theoretic considerations apply, it is prudent to distinguish on the one hand between the
physical vacuum, which possesses some properties relevant for signal propagation;
and, on the other hand, space-time frames, which are constructions based on and ``tied to'' some idealized physical properties of vacuum.
One such typical assumption entering the formal derivation of the transformation properties of inertial space time frames
is the constancy of the velocity of light in vacuum, regardless of the state of inertial motion of any observer.

\begin{acknowledgments}
This research has been partly supported by FP7-PEOPLE-2010-IRSES-269151-RANPHYS.
\end{acknowledgments}


%

\end{document}